# Effect of unbalanced and common losses in quantum photonic integrated circuits

Ming Li(李 明)[1,2], Chang-Ling Zou(邹长铃)[1,2], Guang-Can Guo(郭光灿)[1, 2], Xi-Feng Ren(任希锋)[1, 2,*]

[1] *Key Laboratory of Quantum Information, CAS, University of Science and Technology of China, Hefei, Anhui 230026, China*
[2] *Synergetic Innovation Center of Quantum Information & Quantum Physics, University of Science and Technology of China, Hefei, Anhui 230026, China*
*\*Corresponding author: renxf@ustc.edu.cn*

## Abstract

Loss is inevitable for the optical system due to absorption of materials, scattering caused by the defects and surface roughness. In quantum optical circuits, the loss can not only reduce the intensity of signal, but also affect the performance of quantum operations. In this work, we divide losses into unbalanced linear loss and shared common loss, and provide a detailed analysis on how loss affects the integrated linear optical quantum gates. It is found that the orthogonality of eigenmodes and the unitary phase relation of the coupled waveguide modes are destroyed by the loss. As a result, the fidelity of single- and two-qubit operations decrease significant as the shared loss becomes comparable to the coupling strength. Our results are important for the investigation of large-scale photonic integrated quantum information process.

Photonic integrated circuits [1] (PIC) have been developed for the increasing complexity of both classical and quantum information processing, which is demanding on scalability, stability and high quality interference. By integrating the waveguides and controlling their coupling on a chip, basic optical elements [2] in bulk optics can be realized on-chip with high quality, such as beam splitter (BS), phase shifter and polarization beam splitter (PBS) [3,4]. Recently, quantum C-NOT gate, quantum walk and Boson sampling have been performed on a single chip, based on silica-on-silicon waveguides [5,6], laser direct writing waveguides [7,8] and plasmonic waveguides [9,10]. While, there still remains challenges to integrate optical devices with good performance, and the errors due to experimental imperfection will be amplified when cascading many basic integrated devices together for future quantum computing, simulation and communication.

Among various imperfections, loss is inevitable which is generated from both the essential absorption of materials and the technical problems in fabrication. The effect of loss in bulk optics has been studied in early years [11,12]. When dealing with integrated circuits, many basic optical components are integrated together, and more complex structures should attract our attention. Generally, there are off-chip insertion loss and on-chip waveguide loss. Usually, people summarize all these linear losses and combine them with the inefficiency of detectors. Since quantum processes can be realized via post-selection, which claims successful when detecting the photons in the desired manner, so linear quantum computation can still be performed with those imperfections, and the only influence is the low success probability.

In this paper, we studied the general loss model in the on-chip beam splitter (BS) devices and its effects on the gate fidelities. We found that when there is unbalanced loss or shared common loss channel in the BS, there will be significant errors that will affect the performance of the optical quantum processing.

For an ideal linear process supported by a quantum PIC, the relation between the input and output field can be described by a unitary matrix. It has been demonstrated that any unitary matrix can be decomposed to the product of two level matrix [13], which meanwhile can be further decomposed to phase shifters and BS [14]. Fig. 1(a) shows a sketch picture of a directional coupler, the physical realization of BS, where two waveguides approach each other and exchange energy. For simplicity, we only focus on the uniform coupling regime, whose properties can be analyzed by solving the eigenmodes of the coupled waveguide at the cross-section. In the weak coupling regime, two waveguides couple with each other through tunneling, which can be quantitatively described by the coupling rate C. According to the coupled mode theory, the dynamics of photon amplitude $A_{1,2}$ in two waveguides should obey

$$\frac{d}{dL}A_1 = (-i\beta - \gamma_1)A_1 - iCA_2 \quad (1)$$

$$\frac{d}{dL}A_2 = (-i\beta - \gamma_2)A_2 - iCA_1 \quad (2)$$

where $A_i$ is the photon amplitude in each waveguide, $\beta$ is the propagation constant, $\gamma_i$ is the damping rate, L is distance along the propagation direction. The coupled equations can also be expressed in vector form

$$\frac{d}{dL}\vec{A} = -i\boldsymbol{H}\vec{A} \quad (3)$$

with

$$\vec{A} = \begin{pmatrix} A_1 \\ A_2 \end{pmatrix}, \boldsymbol{H} = \begin{pmatrix} \beta - i\gamma_1 & C \\ C & \beta - i\gamma_2 \end{pmatrix} \quad (4)$$

Therefore, the eigenmodes can be solved by solving the eigenvector of $\boldsymbol{H}$ as a superposition of the field in two waveguides. If a photon is loaded to one of the waveguide corresponding to the superposition of the two eigenmodes, and the photon will oscillate between the waveguides in the sine function form since the eigenvalues of the two eigenmodes are different.

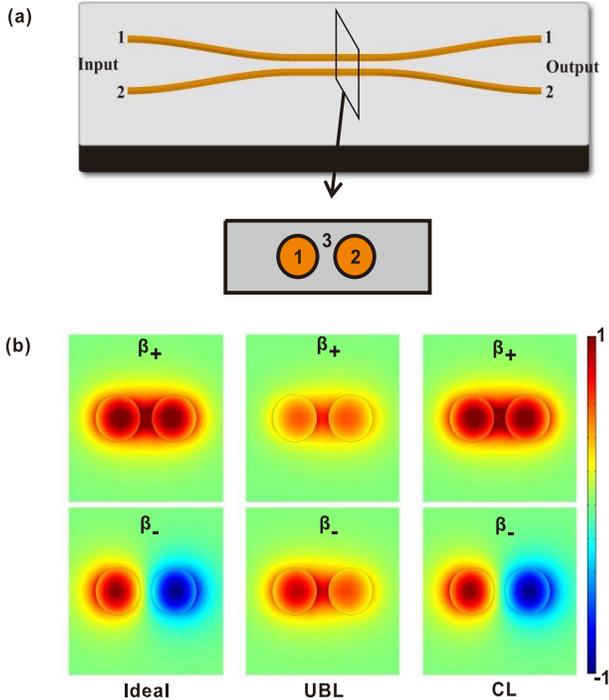

Fig. 1. Eigenmodes of coupled waveguides. (a) Schematic of directional coupler. (b) The cross-section electric field distributions of the eigenmodes for different case. From left column to right, the pictures correspond to the (i) ideal case $\gamma_1 = \gamma_2$, (ii) $\gamma_1 \neq \gamma_2$ and material 3 is lossless, (iii) $\gamma_1 = \gamma_2$ and material 3 is absorptive.. For the left and right case, the two eigenmodes are orthogonal and the overlap is zero. For the middle case, the overlap of the two modes is positive.

In general, we assume the two waveguides and their surrounding are made by different materials, and the configurable structures can be divided into three cases: (i) two waveguide are identical, thus the propagation losses $\gamma_1 = \gamma_2$, (ii) the two waveguides are made by different materials ($\gamma_1 \neq \gamma_2$) and the surrounding material is lossless (iii) $\gamma_1 = \gamma_2$ while the surrounding material is absorptive. For case (i), the propagation loss of light in the two waveguides can be taken out as a global damping factor, thus the two waveguides couple with each other as they are ideal, except the attenuation of the amplitude. The left column in Fig. 1(b) shows the eigenmodes for case (i), the two modes are orthogonal as the field in two waveguide are in-phase and out-of-phase, respectively, while the amplitude in two waveguides are identical. When dealing with hybrid PICs, the case (ii) should be considered. The modes in two waveguides attenuates with different rate and exchange energy continuously with each other when they propagate forward. However, the different damping rate make them no longer orthogonal, which is explicitly shown in the middle column of Fig. 2(b). In all the cases, the two eigenmodes have different energy distributions. If materials 3 are absorptive, different energy penetration to material 3 means different damping loss, corresponding to case (iii). For long coupling length, the eigenmode with bigger loss becomes negligible comparing with the one with smaller loss. Even through the two modes are still orthogonal, the different amplitudes destroy the interference. For cases (ii) and (iii), the imperfections are named as unbalanced loss (UBL) and common loss (CL), respectively. From the preliminary analysis, we can judge that loss can indeed change the evolution of photonic quantum states in the PIC.

For case (ii), the UBL usually exists in hybrid coupled waveguides, for example two waveguides supported by dielectric and metal materials [15,16]. Besides, the UBL may appear if the fabrication roughness or curvature of two waveguides are different.

By Eq. (4), the outputs of two waveguide can be solved as $\vec{A}(L) = e^{\int_0^L -iH dL} \vec{A}(0) = \mathbf{M}(L)\vec{A}(0)$ with L being the length of the coupling region. We obtain

$$\mathbf{M}(L) = e^{(-i\beta - \frac{\gamma_1+\gamma_2}{2})L} \times \begin{pmatrix} \frac{\Delta\gamma}{2\alpha} Cosh(\alpha L) - Sinh(\alpha L) & \frac{iC}{\alpha} Sinh(\alpha L) \\ \frac{iC}{\alpha} Sinh(\alpha L) & \frac{\Delta\gamma}{2\alpha} Sinh(\alpha L) + Cosh(\alpha L) \end{pmatrix}$$

(5)

where $\Delta\gamma = \gamma_1 - \gamma_2$ and $\alpha = \sqrt{\frac{\Delta\gamma^2}{4} - C^2}$. For the bosonic operator of input ports $a_{1,2}^{in}$, the output should be ($j = 1,2$)

$$a_j^{out} = [\mathbf{M}(L)]_{j,1} a_1^{in} + [\mathbf{M}(L)]_{j,2} a_2^{in} + \sum_0^\infty p_k c_k \quad (6)$$

where $c_k$ is the accessible environment mode and $p_k$ is the amplitude probability that satisfying $\sum_k |p_k|^2 + |[\mathbf{M}(L)]_{j,1}|^2 + |[\mathbf{M}(L)]_{j,2}|^2 = 1$.

In Fig. 2(a), we plot the probabilities $|[\mathbf{M}(L)]_{j,1}|^2$ as a function of L. These two curves still behave sine function oscillation accompanied by exponential decay. The inset shows the oscillation curves by eliminating the global damping factor $e^{-\frac{\gamma_1+\gamma_2}{2}L}$. We find an unusual phenomena that the phase between the two sine oscillation is no longer π, which is different from the ideal case. To explain the decreased phase, we investigated the change of the eigenmodes altered by such unbalanced loss. By diagonalizing matrix $\mathbf{H}$, we get the eigenvalues $\beta_+$, $\beta_-$ and eigenvectors $|+\rangle$, $|-\rangle$, corresponding to the effective mode propagation index and the wave function of the eigenmodes, respectively. The eigenvalues are,

$$\beta_{\pm} = \beta - \frac{i(\gamma_1+\gamma_2)}{2} \pm \sqrt{4C^2 - \Delta\gamma^2} \quad (7)$$

First, we investigated the mode orthogonality of the eigenmodes, and plot the overlap of the two eigenmodes $|\langle+|-\rangle|^2$ in Fig. 2(b). The unbalanced loss destroyed the orthogonality of the eigenmodes. As the unbalanced loss $\Delta\gamma$ approaches C, the overlap between the two modes increases to unity, which means the two modes almost identical. As a result, the phase between the two oscillations in Fig. 1(a) decreases from π to 0 and two curves get closer. The broken of mode orthogonality was also demonstrated by the change of effective mode index of the eigenmodes. Fig. 2(c) gives the real and imaginary part of the difference of $\beta_+$ and $\beta_-$. $\Delta\beta$ decreases as $\Delta\gamma$ changes close to 2C . It should be noted that, at the critical point $\alpha = 0$, two eigenmodes becomes totally the same, which is called the exceptional point [17,18]. The two eigenmodes completely overlap and share the same mode index. At the exceptional point, the two-dimension system coalescence to one dimension. By diagonalizing the

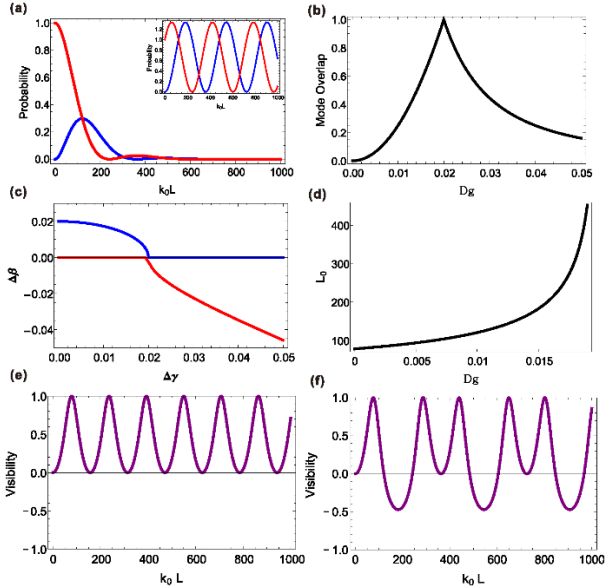

Fig. 2. Two coupled waveguides with unbalanced linear loss. (a) By injecting single photon in port 1 in Fig. 1(a), the hopping probability to the outputs are plotted. Blue line: port 1. Red line: port 2. The inset shows the relative intensity eliminating the global damping factor. We set $C = 0.01k_0$ and $\Delta\gamma = 0.01k_0$ . (b) The mode overlap $|\langle+|-\rangle|^2$ of the two eigenmodes. (c) The real (blue) and imaginary (red) part of the difference between $\beta_+$ and $\beta_-$ . The cross point is the exceptional point. (d) $L_0$ is the minimum coupling length to achieve 1:1 splitting. (e). Two photon quantum interference visibility on an ideal BS with $C = 0.01k_0$ . (f) Two photon quantum interference visibility on a BS with unbalanced linear loss as a function of the coupling length. $C = 0.01k_0$ and $\Delta\gamma = 0.01k_0$ . In all the figures, the units of C and $\gamma_i$ are the free space wave vector $k_0$ . We set C=0.01 $k_0$ for all the cases.

Hamiltonian of the coupled system, we find that there is only one eigenvector $(i,1)^T$ . In an isolated Hermitian system, coupled modes or energy levels will repel each other and the levels will never cross, which is called anti-crossing. While in such a non-Hermitian system, the two levels coalesce to one level, although they are not degenerate and the two dimension system is reduced to one [18], which has been observed in coupled micro-cavities [19] and coupled waveguides [20] with loss-gain. Here we show that coupled waveguides with only loss also have this phenomenon.

Another important change induced by the UBL is the increase of the oscillation periods of the intensity lines, compared to the ideal case. In Fig. 2(c), the increase of the oscillation periods means the coupling strength between the waveguides has been weakened by the loss difference $\Delta\gamma$. For example, to get a 1:1 BS, longer coupling length is required. Fig. 2(d) gives the relation between the minimum coupling length $L_0$ for 1:1 BS and the unbalanced loss $\Delta\gamma$. We can conclude that the effective coupling strength is weakened by the unbalanced linear independent loss. Note that $\gamma_1 = \gamma_2$ is the zero point in the curves in Fig. 2, which behaves the same with the ideal case, except for the decay of the total energy.

In quantum optics, the interference of indistinguishable photons in a waveguide circuit is very sensitive to the phase relation between different input and output ports, which reflects in the phase difference between different elements of the process matrix **M**(L) . One basic process to test the quantum nature of the circuit is HOM interference [21]. By injecting two indistinguishable photons from the inputs, we can calculate the second order quantum correlation $\Gamma_q$ of the output state. Comparing the quantum correlation with classical correlation $\Gamma_c$ , the contrast can be used to measure the performance of the BS for quantum operation. Here, we use visibility V to characterize the performance of BS in quantum optics,

$$V = 1 - \Gamma_q / \Gamma_c. \quad (8)$$

From Eq. 6, we can calculate the second order classical and quantum correlation $\langle a_1^{out\dagger} a_1^{out} a_2^{out\dagger} a_1^{out} \rangle$ of

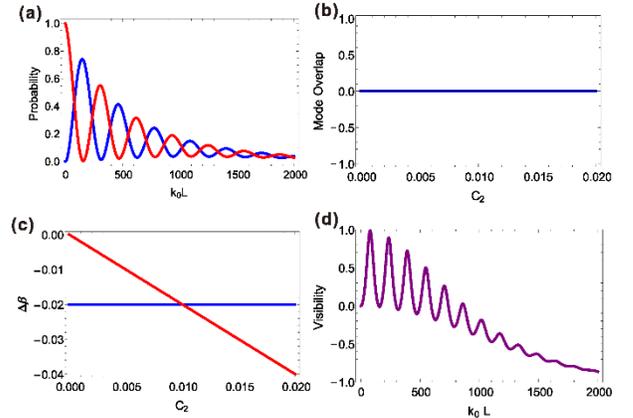

Fig. 3. Single photon and two photon interference on a BS with shared common loss. (a) Relative probability in two waveguides with single photon input. (b) Mode overlap $|\langle+|-\rangle|^2$ of the two eigenmodes. (c) Real and imaginary part of $\beta_+ - \beta_-$. (d) Visibility of two photon quantum interference. The visibility becomes negative and approaches -1 as the coupling region becomes longer. Here we set the damping rate the waveguides $\gamma = 0.001k_0$, $C_1 = 0.01k_0$ and $C_2 = 0.0005k_0$.

the output state as

$$\Gamma_c = |M_{11}M_{22}|^2 + |M_{12}M_{21}|^2, \quad (9)$$
$$\Gamma_q = |M_{11}M_{22} + M_{12}M_{21}|^2 . \quad (10)$$

As is shown in Fig. 2(e), the visibility of an ideal coupler oscillates between 0 and 1 periodically, with the increasing of coupling length L of the BS. However, in the case with UBL, the adjusted phase relation makes the quantum interference different. The relation between visibility and coupling length is plotted in Fig. 2(f). Two main obvious difference can be found, 1) the period of oscillation becomes large, which is consistent with the result of single photon analysis, 2) the visibility appears to be negative in some coupling region. This result is totally different from the HOM interference on an ideal BS, in which case, the second order quantum correlation is always smaller than the classical correlation. As the loss difference $\Delta\gamma$ approaches the exceptional point, the second order quantum correlation becomes twice the classical correlation, thus the visibility gets its minimum value -1. A HOM dip can never be observed at the point, and a peak is observed instead. The reason is that loss destroyed the orthogonality of the non-Hermite system, so quantum coherence cannot be fully maintained in this devices.

From the analysis for single and two photon state, we can conclude that the UBL not only influences the efficiency of the integrated circuit but also changes the function of the circuit. In one hand, the effective coupling strength is weakened by the loss difference. As a result, larger circuit should be designed to realize the same operation. On the other hand, the impacted phase relation leads to extraordinary interference for both classical and quantum field. For waveguides with unbalanced losses, the energy cannot be exchanged between the two waveguide with complete coherence. So, when designing integrated devices for quantum operation, the loss difference between different waveguides should be carefully controlled, especially for waveguides with very weak coupling strength.

In the above paragraph, we discussed the effect of the UBL, while the coupling rate C between the waveguides is still real. However, there exists some cases that the coupling constant $C$ to be a complex number because the two coupled waveguides not only couple in the regime of the direct overlap of their modes, but also through the reservoir in surrounding material, such as in plasmonic circuit. To study such effect, we add an imaginary part $iC_2$ to the real coupling constant $C_1$. Similar to the procedure in the UBL case, we calculate the transfer matrix elements in the CL case, where the probabilities still oscillate in the form of sine function with opposite phase (shown in Fig. 3(a)). With an increasing L, the oscillations get less pronounced. As the coupling length becomes large enough, the two curves tends to merge to a single line and the splitting ratio of the BS approaches 1:1. Fig. 3(b) and 3(c) show the properties of the eigenmodes of the CL case. Fig. 3(b) indicates that the two eigenmodes are still orthogonal. The altered effective mode index are plot in Fig. 3(c).

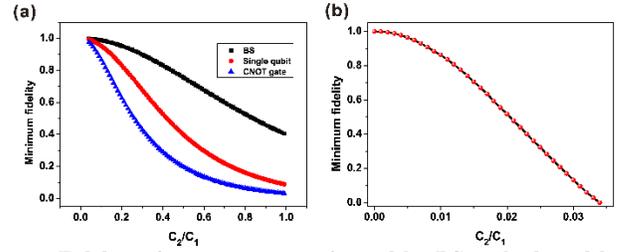

Fig. 4. Fidelity of quantum gates formed by BS with shared loss. All quantum gates are decomposed to BS and phase shifters and we assume the phase shifters are ideal. The fidelity is the minimum value searched through all input quantum states. a. The gate fidelity for BS, single qubit operation and quantum C-NOT gate. b. Minimum fidelity for any two-qubit gate and any two-qubit quantum state. In the calculations, $C_1 = 0.01 k_0$.

We then examine the quantum performance of the coupled two waveguides with CL by calculating the visibility. As is shown in Fig. 3(d), the V of two-photon interference decreases from 1 to -1, with the coupling length increasing to infinity. This arises from the reason that the two eigenmodes of the coupled system decrease with different rate, and the system acts as a filter. When L becomes large enough, the mode with bigger loss can be neglected compared to the other one, thus any photon will be found with equal probability in each waveguide. To explain the visibility of -1, we consider the non-Hermitian system as a subsystem of a higher dimensional Hermitian system, where the process matrix **M** is only a submatrix of a unitary process by including all environment degree of freedoms (Eq. 6). The bunching nature of the photons causes $\Gamma_q = 2\,\Gamma_c$ when L becomes infinite large, which leads V to be -1. Under this situation, V can be treated as a feature of the degree of non-unitary of the process matrix. The CL loss is responsible for the low quantum interference visibility in plasmonic circuits [22], which can be avoided using dielectric loaded circuit [9]. Recently, HOM interference showing peak fringe was experimentally observed in the plasmonic system [23].

We then go further to investigate how such loss influence the quantum operation of a relatively large circuit formed by BSs and phase shifters. On a real integrated optical chip, the shared loss appears to be more influential. From simple to complex, we calculated the fidelity of single qubit operation, C-NOT gate and arbitrary two qubit gate suffered from shared loss. Following the quantum gate decomposition method, these gates are decomposed to BS and phase shifters. Here, we replace the ideal BSs with BSs suffering from shared loss. The fidelity F of the gate is defined as

$$F = min_{|\Phi\rangle, U}\sqrt{\langle\Phi|U^{\dagger}B|\Phi\rangle\langle\Phi|B^{\dagger}U|\Phi\rangle} \qquad (11)$$

where $|\Phi\rangle$ is the input state, U is the unitary operation of the ideal circuits, B is the real quantum operation using BS with loss. The fidelity $F$ is searched among all the input states and quantum operations. Here, we change the relative value of $C_2$ to evaluate the fidelity of quantum gates. From Fig. 4(a) and 4(b), we see that, as the imaginary part of $C$ gets comparable to its real part, the fidelity decreases fast. Comparison between quantum gates shown in Fig. 4(a) and 4(b) also indicates the circuit complexity is very sensitive to the imaginary part of the coupling rate. To achieve fidelity of two qubit gate higher than 99%, $C_2 < 2.5 \times 10^{-4} C_1$ should be ensured. For larger quantum circuits, the dependence on $C_2$ will be more sensitive. So when dealing with large-scale circuits supported by absorptive materials, reasonable design and quantum error correction should attract much more attention to avoid such shortcomings.

In summary, we have investigated the performance of a realistic coupled waveguides system, which is the basic element in quantum PIC. As an inevitable factor in reality, loss is a block in the way of integrating optics on chip in large scale. Apart from its harmful effect on the efficiency, loss also change the physical process. By analyzing the property of the eigenmodes and evaluating the second order coherence of two-photon quantum interference, we quantify the performance of the coupled system. We find that unbalanced linear loss can weaken the effective coupling strength and destroy the orthogonality of the coupled waveguides. Both the unbalanced linear loss and the complex coupling between two identical waveguides result in a reduction of the visibility of the quantum interference. All these imperfections should be considered when designing and fabricating large-scale quantum PICs.

This work is supported by National Natural Science Foundation of China (Grant No. 11374289, 61590932, 61505195), National Key R & D program (Grant No.2016YFA0301700, 2016YFA0301300), the Innovation Funds from the Chinese Academy of Sciences (Grant No.60921091), the Fundamental Research Funds for the Central Universities and the Open Fund of the State Key Laboratory on Integrated Optoelectronics (IOSKL2015KF12). We thank Xiao Xiongfor useful discussion.